\documentclass[article]{svmono}

\usepackage[english]{babel}
\usepackage{marvosym}
\usepackage{graphicx}        

\def\3H{{3\over2}}
\def\simlt{\lower.5ex\hbox{$\; \buildrel < \over \sim \;$}}
\def\simgt{\lower.5ex\hbox{$\; \buildrel > \over \sim \;$}}

\def\simlt{\lower.5ex\hbox{$\; \buildrel < \over \sim \;$}}
\def\simgt{\lower.5ex\hbox{$\; \buildrel > \over \sim \;$}}

\begin{document}



\pagestyle{plain}
\ \ 
\vspace{1cm}

\begin{center}
  {\huge {\bf Karl Schwarzschild, Annie J. Cannon}}
  \bigskip
  
  {\huge {\bf and Cornelis Easton:}}
\bigskip

{\huge{\bf PhDs honoris causa of Jacobus C. Kapteyn}}
\vspace{1cm}

\noindent
{\LARGE Pieter C. van der Kruit},\\
{\Large Kapteyn Astronomical Institute, University of Groningen,}\\
{\Large P.O. Box 800, 9700AV Groningen, the Netherlands}.\\
{\Large email: vdkruit@astro.rug.nl}
\vspace{2cm}

\end{center}

{\large

\noindent
\vspace{2cm}

\abstract{Honorary degrees and particularly doctorates are important
instruments to enhance the standing of universities and professors, in
addition to receiving these as a measure of a scientist’s recognition. Jacobus
C. Kapteyn from the University of Groningen in the Netherlands, one of the
most prominent astronomers of his times, received three of these and has
persuaded his university to award at least three, possibly five. I examine the
background of the selection of the latter in view of developments in Kapteyn’s
times in his career, international astronomy and political  and cultural
circumstances.}
\bigskip

\newpage

\section{ Introduction}

Universities boost their standing and prestige by appointing or
keeping important scientists
on their staff that have performed groundbreaking research, preferably
while being employed by them, but if necessary when performed elsewhere.
Listing alumni that have become leading researchers anywhere else but who
started their careers by defending their thesis work at the university involved
is also used to justify claims concerning success and  status. Another
important instrument is awarding honorary doctorates to prominent scientists
or cultural or political figures.

One aspect
of a professor's success may be taken to be who his or her students are.
This involves for professors at most universities supervising research
of students that leads to the defense of an important thesis, after which
conferring a doctorate on the student as `promotor'\footnote{In the
Netherlands, Belgium, Germany and a few other countries the professor who has
supervised the research is designated as the promotor and bestows the degree
upon the candidate on behalf of the university. When I use this word `promotor'
below it may also refer to the equivalent in other countries.}
on behalf of the Senate (or
other body representing the university). But it is also possible to nominate
persons for the degree of doctor {\it honoris causa} (or for short {\it h.c.})
and acting as a honorary promotor.

Jacobus C. Kapteyn was a professor of astronomy at the
University of Groningen in the Netherlands
from 1878 to 1921. In his days he was one of the most
prominent astronomers in the world, which can be illustrated by the fact that
the most important observatories signed up to his {\it Plan of Selected Areas}
in which observations of stars in each of the 206
{\it Areas} were collected. Harvard
College Observatory, under the directorship of Edward C. Pickering,
photographed all these areas and sent large stacks of
plates to Groningen for measurement
of positions and magnitudes. Even more telling, George E. Hale in 1908 adopted
Kapteyn's {\it Plan} as the primary program for his brand new, giant 60-inch
telescope on Mount Wilson, the largest in the world,  and had Kapteyn
appointed for life as research associate by the Carnegie Institution of
Washington, and between 1908 and 1914 Kapteyn annually visited Mount Wilson.

Kapteyn did not have many students defending their theses under him
during the tenure of his professorship. There were only eight of which
Willem de Sitter undoubtedly is the most prominent. Pieter J. van Rhijn, who
was to become his successor, has also gained some international standing.
Actually Adriaan van Maanen might also be counted as a student of Kapteyn,
since he performed a significant part of his thesis research in Groningen with
Kapteyn; in spite of graduating under Albertus A. Nijland in Utrecht he
regarded Kapteyn as his real mentor. Three more students started their thesis
work while Kapteyn was in office but completed their theses only when van Rhijn
had taken over (and in fact after Kapteyn had died). Of these Jan Schilt
pursued a very successful career in the USA, but without dispute the most
prominent of Kapteyn's {\it nachwuchs}
(i.e. including the ones that started their
research under Kapteyn even though van Rhijn eventually was their thesis
supervisor) is Jan Hendrik Oort. Much more on
Kapteyn and Oort can be found in my academic \cite{JCKbiog}\cite{JHObiog}
and wider-audience \cite{JCKEng}\cite{JHOEng} biographies.

It is not well documented that Kapteyn was promotor in at least three cases of a
doctorate {\it honoris causa}. The first one, awarded to journalist and amateur
astronomer Cornelis Easton, is
usually noted, but those of very prominent astronomers Karl Schwarzschild and
Annie Jump Cannon are much less known. Schwarzschild's is rarely mentioned,
Cannon's is usually listed but understandably eclipsed by the more
prestigious one she obtained  a few years later from Oxford and sometimes only
the latter one is mentioned. 

The circumstances applying to the conferring of these
degrees are the subject of this paper. Questions addressed are the
following. Was this unusual?
Why did Kapteyn select these people? Did Schwarzschild and Cannon
actually come to Groningen to receive the honors? Kapteyn
himself had three doctorates {\it h.c.}, Capetown, Harvard and Edinburgh.
How does this compare with others? He had none from English (distinct from
British), German or French universities, where after all many of
the leading astronomers in Europe were working. Does this tell us something or
doesn't it?

This article is aimed at an audience of historians and others interested, not
necessarily with an astronomical expertise. I therefore will describe
in some detail the required astronomical background to appreciate the
significance of matters discussed. The reader is assumed to be familiar
with basic astronomical concepts.

\section{ Background}

\subsection{\Large Honorary degrees: General}

According to Merriam-Webster an honorary degree is `a degree
given by a college or university to someone who is not a student but who has
done something important' \cite{MerWebs}.

The first such degree seems to have been awarded by the University of Oxford,
UK. On their Website we read \cite{Oxford}:
\begin{quotation}
{\large `The earliest honorary degree (in
the sense which we would understand it today) appears to have been offered to
Lionel Woodville in 1478 or 1479. Woodville, Dean of Exeter and the
brother-in-law of Edward IV, appears to have already held the degree of
Bachelor of Canon Law; the University offered to confer the degree of Doctor
of Canon Law on him without the usual academic exercises. It was thus an offer
to dispense with the usual requirements, but was apparently unsolicited and
clearly an attempt to honour and obtain the favour of a man with great
influence.'}
\end{quotation}

In the USA the first honorary degree has been awarded by Harvard University in
1692 to Increase Mather, a minister and staunch Puritan, who between 1685 and
1791 was Acting-President, Rector and President of Harvard College. Harvard
University claims to have awarded over 2,300 honorary degrees by now
\cite{Harvard}. In the UK the universities of Oxford and Cambridge are
prolific awarding institutions; according to Heffernan \&\ J\"ons \cite{HF07},
these awarded respectively 1,487 and 1,111 doctorates
during the twentieth century. Both universities now award between eight and
ten degrees per year. Harvard University nowadays
awards between 5 and 15 degrees. But these of course are  it not all
doctor degrees. 

The University of Groningen has awarded almost 300 honorary doctorates,
the first occurring in 1717 \cite{RUGhc}, which was for Abrahamus Trommius
(Abraham Trom) in theology for a concordance of both the Old and the New
Testament. However, already from 1618 onward
there were doctorates awarded {\it sine
examine}, so without submitting and defending a thesis. An
interesting example is one awarded in 1634 to famous theologian
Gisbertus Voetius, who was a preacher that was being appointed professor
of theology in Utrecht. It was felt undesirable that he would award PhDs
without having obtained this degree himself and
this is how that was solved. Astronomer Frederik Kaiser in Leiden was a
comparable case; he was promoted {\it honoris causa} in 1835 so that he could be
appointed lecturer in 1837 and eventually in 1840 professor of astronomy.

Honorary doctorates in Groningen
are rare in most years but do come in a larger numbers in years of an
anniversary that is a multiple of fifty years. The peak at the tricentennial
in 1914 was enormous, 67 in total starting with Queen Wilhelmina of the
Netherlands. At other times the university
has been more modest, in 2014 at the fourth centennial (aptly named
`for infinity' or `4{\large $\infty$}') there were only nine.
In recent years that are not a lustrum
(quinquennium) it is mostly zero, sometimes one or exceptionally two.
Groningen doctorates {\it honoris causa} are much, much rarer
than Cambridge, Oxford or Harvard, but the latter ones are certainly more
sought after.

For completeness I note that as a matter of principle some universities do
not confer degrees {\it h.c.} at all, such as Cornell University, the
Massachusetts Institute of Technology (MIT) or Stanford University. All Dutch
universities do award honorary degrees.

Like the first one in Oxford, doctorates {\it h.c.} are awarded
often to royalty and political figures. In case of the Netherlands, for example
Queen Juliana of the Netherlands has received honorary doctorates from both
Leiden and Groningen and in addition eight more from foreign universities. Some
renowned persons have tens of such honorary degrees; for example Nelson
Mandela had over fifty, in the Netherlands from Leiden  university.
Kapteyn's University of Groningen honored the Queens of the Netherlands,
Wilhelmina in 1914 at its 300-th anniversary and Juliana at its 350-th, and
more recently international dignitaries Helmuth Kohl, Desmond Tutu and Ban
Ki-Moon. This may sometimes be controversial, e.g. for the case of Groningen
when Kohl a few years later
was implicated in the CDU donations scandal, or when Oxford University's
governing assembly, Congregation, by a large majority
refused Margaret Thatcher an honorary doctorate. 

\subsection{\Large Honorary degrees: University of Groningen}

For reference to the discussion in the remainder of this article, I give some
statistics on doctorates {\it honoris causa} in Groningen during Kapteyn’s
tenure of his professorship, actually between 1878 when he was appointed and
1922, one year after his retirement and the time of his death. I already
alluded to the remarkable peak of 67 in the year of the university’s
tricentennial, 1914. There were 14 in Kapteyn's day before that year and 5
after. The first one in 1884 did not set the scene; it was famous physician
and microbiologist H.H. Robert Koch, one of the founders of
bacteriology and the concept of infectious disease.
He would receive the Nobel Prize in 1905 for identifying the bacterium that
caused tuberculosis. But after that there
were only Dutch persons that had important academic contributions on their
record, but for one reason or the other had never been in a position to
present a PhD thesis; this is of the kind of what we might call `corrective'
honorary doctorates. One of these was amateur astronomer
Cornelis Easton, to whom I will return below. One other example was the first
woman that received a honorary doctorate in Groningen,  Jantina Tammes,
who studied for and obtained secondary teacher qualifications in a number of
natural sciences and came to work as assistant of Willem Moll, professor of
botany in Groningen. Her contributions led to the award of a doctorate
{\it h.c.} in 1911. She went on to become the first female professor in
Groningen and the second in the country (Johanna Westerdijk, also botanist, in
Am\-ster\-dam
preceded her; it is telling for the slow progress of women equality
that both were extraordinary, that is  professorships added to the ordinary
contingent for the individual involved). Except for Koch in 1884 all
up to the 1914 peak are of this type of `corrective' honorary doctorates.

Another good example is Maria H.J.P. Thomassen, who was promoted in 1905
{\it honoris causa} in the Faculty of Medicine. This is
an amusing case for another
reason, namely that this person was incorrectly claimed on the
basis of the list on the University's Website \cite{RUGhc}
to be a physician and the first female
doctor {\it h.c.} in Groningen \cite{ndhnbm}. However Maria Thomassen was
definitely male (but from Roman Catholic parents in the
Dutch southern province of
Limburg, where boys named Maria were not uncommon) and not a physician,
but a veterinarian. He had studied at the prominent School for Veterinarians
in Utrecht, which could not award doctorates. Later he became a teacher at this
school and did significant scientific research, some of which together with
Hartog Hamburger, professor of physiology in Groningen, who acted eventually as
honorary promotor. This School in 1925 would become a Faculty within the
University of Utrecht and then doctorates could be awarded.

It should be noted that cases of `corrective' honorary degrees were not
uncommon, because the occasion arose easily.
An important illustration is related to the admission to universities. This
was at first restricted to boys (no girls yet) from the Gymnasium (grammar
school). But in 1863 a new school type was
introduced (Hoogere BurgerSchool HBS or Higher Civic School),
in which much attention was paid
to mathematics and natural sciences, but where no Greek and Latin was taught.
This was a preferred route around the Fin de Si\`ecle (the end of the
nineteenth century) for boys and girls
attracted to mathematics and science, where admission to the
university was accomplished by taking a special, additional exam. It is the
route many Dutch Nobel Prize winners of that period had followed. Also
astronomer Jan Hendrik Oort did this. Now persons having gone through the HBS
were like others admitted to defend
a thesis, except for a long time in Medicine. Oort's father was a physician
specializing in psychiatry and, having attended the gymnasium,
had defended a thesis in Medicine in Leiden under the first
professor of psychiatry in the Netherlands, Gerbrandus Jelgersma. Jelgersma
had attended the HBS and then studied medicine, and for this
reason was not allowed to submit a thesis. This was solved by the University
of Utrecht by awarding Jelgersma a doctorate {\it h.c.} Other physicians went
to a university abroad to defend a thesis.
\bigskip

In 1914 there were 44 foreigners among the doctores {\it honoris causa}
in Groningen,
one of which was Karl Schwarzschild. The other 43 were mostly important
scientists, notably L\'eon Duguit, leading French scholar of public law,
Niels Thorkild Rovsing, Danish surgeon,
Arthur Louis Day, American geophysicist and volcanologist,
Henri Pirenne, Belgian historian, and British Alicia Stott (n\'ee
Boole), female mathematician who never held an academic position, but
nevertheless  made a number of valuable contributions to her field. These
were definitely leaders in their field, but probably the most important scientist
was Svante August Arrhenius  from Sweden, founder of physical
chemistry, who had received the Nobel Prize for Chemistry in 1903. 

The remainder were influential Dutchmen, such as famous architect Hendrik
Petrus Berlage, social democrat but known as communist sympathizer,
well-known poet Albert Verwey, and musician, music teacher, composer and
conductor of classical music  Peter Gijsbert van Anrooy, who was
a very good friend of Kapteyn. But also Carel Coenraad Geertsema,
prominent politician and public administrator and at the time also
President-Curator of Groningen University. 
A special case was the doctorate awarded to steel magnate and philanthropist
Andrew Carnegie. The archives of Groningen University contain a telegram from
him in which he profoundly thanks for the honor but notifies that he would not
be present at the celebrations.

In 1914 there were 43 ordinary professors at the University of
Groningen, so most of these could
have proposed two candidates (or more) for these honorary degrees. A number
of them must have been chosen not by individual professors, but
by the governing bodies within the university as
a whole, especially the ones that concerned
politicians or public administrators. Since
van Anrooy was a very good friend of Kapteyn, the latter might very well have
taken the initiative and proposed his name. Carnegie might very well have come
out of Kapteyn's sleeve as well. After all, there was no other connection to
my knowledge between the university of Groningen and Carnegie other than
Kapteyn's appointment by the Carnegie Institution. It might have been a way of
showing gratitute for this and the excellent and unique opportunities offered
to Kapteyn to conduct his research at the Mount Wilson Observatory. Then
it would make sense that it was Kapteyn himself who had come up with the
idea.

Between 1914 and 1922 there were only five more honorary doctorates,
four to Dutch public figures and one to Annie Cannon. Of the 19 awards
during Kapteyn's professorship and outside the celebrations in
1914, only two were international scientists (Koch and Cannon), the rest Dutch
persons that were awarded the degree in the `corrective' sense defined above.
That Kapteyn was allowed to propose a degree for a prominent foreign scientist
in an ordinary year was very unusual and testifies to his exceptionally
prominent status within the university. After all, Kapteyn's fame especially
in the US, evidenced by his collaboration at Mount Wilson and his appointment
by the Carnegie Institution, must not have gone unnoticed by his peers, such
as the other professors and fellow members of the Senate. 
The numbers suggest that having proposed three and having that many
accepted must have been highly unusual, but Kapteyn's status within the
university was such that if someone was allowed that many it would have to be
him.
\bigskip

On the other hand we may ask: which {\it Groningen}
professors were honored with a doctorate {\it h.c.} elsewhere
between 1878 and 1922? This information can be found from the annual
summary of events (the `Lotgevallen' or Happenings, literally `Fates') in the
University by the Rector Magnificus. However, this may not always be complete,
since the relevant installment fails to mention Kapteyn's honorary degree from
Harvard! 
These were:
\begin{quotation}
{\large
$\bullet$ Bernard Hendrik Kornelis Karel van der Wijck, professor of philosophy
and logic (Edinburgh 1884),\\
$\bullet$ Jan Willem van Wijhe, professor of anatomy and embryology, founder of
the Anatomical Laboratory (Freiburg 1889, Aberdeen 1906),\\
$\bullet$ Jacobus Cornelius Kapteyn (Cape Town 1905, Harvard 1909,
Edinburgh 1921),\\
$\bullet$ Hartog Jacob Hamburger, professor of physiology and histology,
founder of the Physiological Laboratory (Aberdeen 1906, Utrecht 1922, Padua
1922),\\
$\bullet$ Anton Gerard van Hamel, clergyman and professor of French language
and literature, editor of the prominent Dutch literary magazine `De Gids'
(Utrecht 1906),\\
$\bullet$ Barend Sijmons, professor of German language and literature (St.
Andrews 1912),\\
$\bullet$ Franz Marius Theodor de Liagre B\"ohl, professor of Hebrew and
Assyrian (Bonn 1915).}
\end{quotation}
Hamburger's degree in Utrecht was actually at the centennial of
the School for Veterinary Science. The total count is then twelve degrees
bestowed upon seven individuals in 45 years. Kapteyns degrees from Cape Town
and Harvard were the only ones from outside Europe.

\subsection{\Large Jacobus Cornelius Kapteyn}

Jacobus C. Kapteyn (see Fig.~{\Large \ref{fig:figJCK}}),
who lived from 1851 to 1922, is
one of the most prominent founders of the field of statistical astronomy.
A general introduction to this field with much backgrouns and Kapteyn's
contributions can be found in \cite{Paul}, to Kapteyn himself see  my
biographies of him \cite{JCKbiog}\cite{JCKERng}.
Kapteyn's  professional life’s aim was to determine the distribution of the
stars in space, building on the approach of the Star Gauges by William Herschel
about a century before him. His first contribution was the provision of a
catalogue of stars in the southern hemisphere comparable to the {\it Bonner
Durchmusterung} in the north and for this his part was to measure out the
photographic plates taken of the complete southern sky by David Gill from Cape
Town. This took him twelve years of labor. The {\it Cape Photographic
Durchmusterung} was published in three installments, the last one
in 1900, the final year  of the nineteenth century,
when Kapteyn approached the age of fifty. He developed the method of
statistically estimating distances of groups of stars, so-called secular
parallaxes, using their proper motions which contain a reflection of the
motion of the Sun through space in addition to the peculiar motion of each
star itself. His assumption was that these latter components
were homogeneous and isotropic. For
his goal of constructing the structure of the sidereal system he needed two more
assumption, namely  that there was no attenuation of light in space by
scattering or absorption by interstellar dust and that everywhere the mix of
stars was the same.

By careful study of proper motions across the sky and looking for systematic
patterns he discovered that this first
assumption was false, and that in addition to
random motions of the stars there seemed to be a set of two systematic
streams, -- when corrected for the motion of the sun --
precisely opposite to each other, accurately directed along the Milky
Way, and, using radial velocities from elsewhere, at a relative speed of about
40 km/sec. These are Kapteyn’s {\it Star Streams}, which he announced at a
large international
congress during the St. Louis World Exhibition in 1904. Karl
Schwarzschild, as we will see, soon came up with an alternative explanation
that proved in the end to be the correct one. Of course the presence
of the {\it Streams} significantly
complicated Kapteyn’s use of secular parallaxes, but did not make it impossible.

In order to proceed with his program to map the stellar distribution in space,
he devised a very ambitious undertaking, defining 220 areas of sky that avoided
bright stars and unusual crowding of fainter stars to derive properties of all
stars to as faint levels as possible: magnitudes, colors, proper motions,
radial velocities (because of the difficulty of obtaining sufficiently
accurate spectra for all but the brighter  stars this was not
restricted to the formal areas), etc. 

\begin{figure}[t]
\sidecaption[t]
\includegraphics[width=0.64\textwidth]{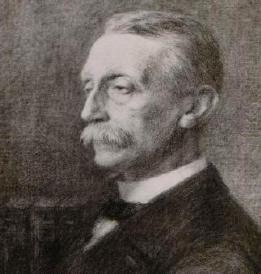}
\caption{ Jacobus C. Kapteyn from a drawing in 1908 by Cornelis Easton. Kapteyn
Astronomical Institute.}
\label{fig:figJCK}
\end{figure}

One very fundamental new property that he needed to consider as well was the
spectral type, a classification of the stars in terms of the lines in their
spectra that proved later fundamental in understanding differences among stars
and their structure and
evolution. This classification scheme was defined around 1901 by the
work of Annie J. Cannon (building also on the work of others), although at the
time the physics and astrophysics underlying it was not at all understood.

This {\it Plan of Selected Areas} was defined in 1906 and Kapteyn succeeded in
assuring commitment of provision of observational material from some twenty
major observatories. In the Introduction I already mentioned Harvard College
Observatory under E. Pickering and Mount Wilson Observatory under George E.
Hale for providing large amounts of observing time to obtain plate material
for the star counts and colors. Some notable further contributions were from
the German Sternwarten of Bergedorf (Hamburg) and Potsdam (Berlin) for spectral
typing using objective prisms, Radcliffe (Oxford), Yerkes, Yale, and the Cape
for proper motions and parallaxes, Lick and Mount Wilson for radial velocities,
and quite a few others. It took decades to complete all of this.

To briefly round up the story on Kapteyn, I note that he did a number of
studies on interstellar absorption (extinction), correctly inferring that it
would be stronger in the blue and therefore give rise to increasing reddening
of stars with increasing distance, but he kept worrying that he
automatically selected in
his studies different mixes and therefore colors of stars with distance. He
actually arrived at quite reasonable values for the amount of extinction, as
we know now, although to a significant
extent this was fortuitous. In 1916, he quickly
accepted Harlow Shapley’s evidence for absence of absorption when the latter
found that with Kapteyn’s absorption stars in globular clusters would have to
be intrinsically some two magnitudes redder than observed.
The option that dust was
restricted to the plane of the Milky Way did not seem to have occurred to
Kapteyn, Shapley or any other influential astronomer at the time. Although
astronomers suspected the  presence of absorbing dust, it took untill the work
by Robert J. Trumpler in 1930 on diameters of star clusters that the case for
interstellar absorption was convincingly settled and it had become clear that it
was restricted to a thin layer in the plane of the Galaxy.

Neglecting absorption Kapteyn near the end of his
life used star counts and proper motions to make a ‘first attempt’ at deriving
a rather flat model of the sidereal system. His final major contribution was to
open up the field of stellar (nowadays usually called galactic) dynamics, where
he explained the equilibrium of the spatial distribution
as a balance between the motions of stars and
their collective  gravitational force. In the perpendicular direction the
distribution was maintained by random motions, in the plane of the system in
addition by rotation and the associated centrifugal motion
which he found was adequately provided by assuming two
systems of stars rotating in opposite directions, which he identified with his
{\it Star Streams}. The sun was in the inner parts, some 650 pc from the center. 

Now, this contrasted with Shapley’s more spherical and larger system of
globular clusters. Eventually this was brought together in the picture of a
disk and halo, respectively Kapteyn’s system expanded by a factor of three or
so when extinction was realized to affect the stellar distribution in the
plane, and Shapley’s shrunk by  factor two. It took until 1938 before Jan
Hendrik Oort used the accumulating data in the {\it Plan of Selected Areas} to
repeat Kapteyn’s analysis allowing for absorption. For detailed discussions
and references to major papers in the development of all of this see
references \cite{JCKbiog} through \cite{JHOEng}. Note that any discussion
on Kapteyn suffers from the fact that his archives, curiously except for the
letters David Gill sent him, have been lost, presumably during the bombing of
Rotterdam in 1940 (see contribution by Petra van der Heijden in the 1999
`Kapteyn Legacy' symposium \cite{Legacy}).

\section{  Dr.h.c. C. Easton and Kapteyn's
25th anniversary as professor in 1903}

Cornelis Easton (see Fig.~{\Large \ref{fig:figEaston}}) lived from 1864 to 1929.
He was primarily a journalist, but in addition
an accomplished amateur astronomer (see \cite{ABEaston} for biographical
notes; this and the English-language article \cite{EvanE} is based on an
extensive biographical article in Dutch by astronomer Johan W.J.A.
Stein SJ of the Specola Vaticana (the Vatican Observatory) \cite{Stein}).

After his secondary education (indeed HBS), he spent a few years studying
various subjects at what is now known as
the Technical University of Delft, after which
he switched to study French and obtained the qualifications for secondary
school teacher in that subject.  But rather than picking up that profession he
took on jobs as a newspaper journalist, eventually as editor (in-chief) of some
daily or monthly newspapers and magazines. He had developed a strong interest
already as a child in astronomy and as a young man started producing drawings
of the Milky Way, some of which he published. He  very early on wondered what
it would look like `from space', that is to say from outside and  from other
directions, speculating there was spiral structure as in Lord Rosse’s famous
drawing of the Whirlpool Nebula. This gave him the idea that the Milky Way had
in fact the structure of a spiral-shaped star cluster. 

So Easton had been intrigued by the observation of spiral nebulae and became
convinced that our Galaxy should not be any different and had to have spiral
structure as well. He therefore derived a model for the Galactic spiral
structure on the basis of the
assumption that the sun is well away from the center and
in areas in the Milky Way that were relatively bright we were
looking along spiral arms and in darker areas in between such arms. In 1900 he
published his ideas in an article in the Astrophysical Journal presenting it
in the title as {\it A new theory of the Milky Way}.
He eventually improved this using photographs
of the Milky Way from many sources giving rise to the representation in
Fig.~{\Large \ref{fig:Eastonmodel}}, taken from another Astrophysical Journal
publication by Easton in 1913 \cite{Easton13}.
The ring at the outer edge is a sketch of the
Milky Way on the sky between latitudes $+10^{\circ}$ and $-10^{\circ}$, and the
inner part the derived structure. It must be said that Kapteyn seemed to have
judged this to be an unjustified over-interpretation.

\begin{figure}[t]
\sidecaption[t]
\includegraphics[width=0.64\textwidth]{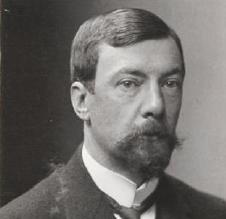}
\caption{ Cornelis Easton in 1906. This picture comes from the
\textit{Album Amicorum}, presented to H.G. van de Sande Bakhuyzen
on the occasion of his retirement as professor of astronomy and director of
the Sterrewacht Leiden in 1908 \cite{LAvdSB}.}
\label{fig:figEaston}
\end{figure}

\begin{figure}[t]
\sidecaption[t]
\includegraphics[width=0.64\textwidth]{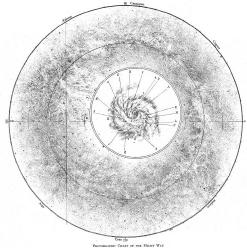}
\caption{ Easton's model for the spiral structure of the Stellar System,
based on a sketch of the Milky Way from many photographs in the outer shell.
Reproduced from Easton's paper in 1913 \cite{Easton13}.}
\label{fig:Eastonmodel}
\end{figure}

Still, Easton’s efforts were sufficient reason to propose a honorary doctorate
in 1903. It is probably not a coincidence that Kapteyn that year celebrated
his 25th anniversary as a professor. In 1878 he was appointed on a new chair in
astronomy in Groningen, spurred by the fact that a new law on higher education
had come into force. This law stipulated among others that the curricula at
the three state-funded universities should be the same and this gave rise to a
very substantial increase in professors at and in the budgets of the
universities. For Groningen a new professorship in astronomy was opened, but
no accompanying observatory, which resulted in Kapteyn's establishment of an
astronomical laboratory, an `observatory without telescopes'. 
So there had been a burst of professorial appointments in 1878
and the Rector Magnificus in Groningen in his annual report on the
`Lotgevallen'  of the University noted with
great satisfaction that of these many jubilees Kapteyn had been the only one
to be honored with a prestigious Knighthood in the Order of the Netherlands
Lion. Maybe the university honored Kapteyn by allowing him to bestow an
honorary doctorate.

From the `Lotgevallen' of the University for the academic year 1902--1903:
\begin{quotation}
{\large `One doctoral award took place {\it honoris causa}, in a public and
extraordinary session of the Senate. On the 13$^{th}$ June Mr. C. Easton from
Rotterdam, where he is an editor of the Nieuwe Rotterdamse Courant, on the
basis of his excellent service for the science of astronomy, was promoted
honorably to doctor in mathematics and astronomy. Promotor was prof. Kapteyn.'}
\end{quotation}

In the NASA/Harvard Astrophysics Data System, Easton has  no fewer than 22
publications, of which 11 in leading international journals including in
addition to the Astrophysical Journal, the Monthly Notices of the Royal
Astronomical Society, Nature, Astronomische Nachrichten, and the Bulletin of the
Astronomical Institutes of the Netherlands. Some of these publications
concern correlations between the distribution of bright stars and the
brightness of the Milky Way and distances of features in the Milky Way.

He was also interested in long term patterns in the weather.  Kapteyn had been
interested in this also. From widths of tree rings in oak trees
in the Trier area in Germany, Kapteyn had inferred  also in the 1880s
a periodicity of 12.4 years
in the amount of rainfall between 1770 and 1880. Easton analyzed `modern'
temperature data for Western Europe since around 1850, less accurate ones in
the hundred years before that and classifications of the harshness of
winters in the centuries before that. Easton then had  found that there was an
89-year period in the occurrence of severe winters, about seven times the 
period of Kapteyn's cycle. Attempts to attribute that
at least partly to the solar cycle were not successful.

Easton eventually became chairman of the amateur organization
Netherlands Association for
Meteorology and Astronomy and editor-in-chief of its periodical Hemel \&\
Dampkring (`Sky \&\ Atmosphere’). For an amateur astronomer his contributions
are remarkable.

\section{ Dr.h.c. K. Schwarzschild and the
University's tricentennial in 1914}

Karl Schwarzschild (see Fig.~{\Large \ref{fig:Schwarzschild}}) lived from
1873 to 1916 (for two authoritative obituaries see Hertzsprung \cite{EH17}
and Eddington \cite{ASEdd}).
Already during his secondary school he published two
papers on determinations of orbits in binary stars and as a student
another paper on variable stars.
He studied under Hugo H. Ritter von Seeliger in M\"unchen (Munich). Von
Seeliger would later produce a model of the sidereal system much
like Kapteyn's and at about the same time. However, it was more
mathematical, schematic and difficult to follow. Schwarzschild
wrote a PhD thesis on Poincar\'e's theory of rotating liquid bodies,
after which he spent some time as assistant at Vienna, where he
developed new methods to derive stellar magnitudes from photographic
plates. This had of course been done before, notably by Kapteyn and Gill
in the {\it Cape Photographic Durchmusterung}
(and also in the {\it Carte du Ciel}), but Schwarzschild
improved the procedure considerably by using extra- or intrafocal
exposures, in the process describing also the characteristic of
reciprocity failure -- although not by that name --
in photographic plates (that at faint levels
it takes more than twice the exposure time to reach the same level
of photographic density for twice as faint incident light). He derived
an empirical formula describing the relation between exposure
(brightness multiplied by exposure time) and the resulting photographic density.
Already in 1901, after a second period in M\"unchen, he was appointed
professor and director of the Sternwarte in G\"ottingen. His most
important work at that time was the
{\it G\"ottingen Actometrie}\footnote{ I
keep the German name \textit{Actinometrie} similarly as I did for
\textit{Durchmusterung} in the CPD.},
a survey of stellar magnitudes between the celestial equator and
$+20^{\circ}$ declination, replacing his method of out-of-focus
exposures by a new method based on a regular shaking of the plate
during exposure to distribute the light over a larger area of emulsion.
The instrument developed
for this was called a {\it Schraffierkassette}. 

\begin{figure}[t]
\sidecaption[t]
\includegraphics[width=0.64\textwidth]{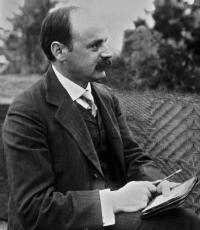}
\caption{ Karl Schwarz-\linebreak[4]
schild in one of the few pictures available of him.
From the English Wikipedia (copyright expired).}
\label{fig:Schwarzschild}
\end{figure}

The {\it G\"ottingen Actinometrie} contained 3500 stars. The word Actinometrie was
derived from the instrument that John Herschel had built, with which he had
determined the energy radiated by the sun by measuring the increase in
temperature in a closed volume that was exposed to sunlight. Herschel had named
that an actinometer, from the Greek word `aktina'
for ray (of light). The term has
also been used by John A. Parkhurst who published about the same time
as Karl Schwarzschild the {\it Yerkes Actinometry} of stars around the north
celestial pole, a kind of precursor of the {\it North Polar Sequence}. The
{\it G\"ottingen Actinometrie} produced an important piece of information,
namely the very tight correlation between stellar colors and spectral type,
thus establishing Karl Schwarzschild's reputation..

In 1909 Schwarzschild was appointed director of
the Astrophysikalisches Observatorium in Potsdam, a very prestigious position.
Potsdam took part in Kapteyn's {\it Plan of Selected Areas},
eventually providing accurate photographic magnitudes for stars in the
northern {\it Areas}
that were too bright for the deep plates taken at Harvard and
Mount Wilson. The Schwarzschild Archives at Nieders\"achsische Staats- und
Universit\"atsbibliothek G\"ottingen has provided me with scans of the relevant
correspondence. Kapteyn and Schwarzschild
corresponded regularly. The Kapteyns visited Potsdam in May 1911
and there is also some correspondence between Mrs. Kapteyn and Mrs.
Schwarzschild, the former writing in English apologizing that German is
`beyond my power'.

Kapteyn and Schwarzschild were on very good terms. The latter
had asked support from Kapteyn to have two  younger astronomers invited to
spend some time at Mount Wilson, for which Kapteyn
mediated by recommending them to George Hale. The first person that Kapteyn
introduced to Hale on recommendation by Schwarzschild
was Ernst A. Kohlsch\"utter, who had obtained
a PhD from Karl Schwarzschild while the latter was still at G\"ottingen and had
moved to Hamburg afterwards. This opened what David DeVorkin has dubbed the
`Pipeline' (see his contribution to the 1999 `Legacy Symposium' on Kapteyn
\cite{Legacy} and that of Klaas van Berkel
in the same volume). Kohlsch\"utter went to
Mount Wilson in 1911 and stayed until the start of WWI. He worked with Walter
S. Adams and they discovered the relation between the width of spectral lines
and absolute magnitude (the dwarf-giant distinction for late types) and
introduced the spectroscopic parallax. Later Kapteyn and Adams had a bitter
argument when Kapteyn felt Kohlsch\"utter did not get the credit he deserved
from Adams.

The second person was Ejnar Hertzsprung, a Danish astronomer working for
Karl Schwarzschild at Potsdam. Kapteyn introduced him to Hale in 1912 and took
him along during his Mount Wilson visit that year. Before that Hertzsprung
spent some time in Groningen and by the time they started out for California
he was engaged to the younger daughter Henriette Kapteyn. They got married in
May 1913 and Henriette moved with him to Potsdam that year.

Although the {\it Actinometrie} was
interesting work to Kapteyn, probably the most relevant in the context of the
honorary degree was Schwarzschild's theoretical work related to the structure
of the Milky Way as a Stellar System. There are two things to mention.

\begin{figure}[t]
\begin{center}
\includegraphics[width=0.98\textwidth]{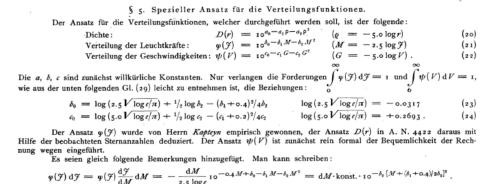}
\end{center}
\caption{ Part of the paper by Karl Schwarzschild in which he developed
a method to solve the integral inversion problem in statistical astronomy.
Here he introduces special analytical forms of the distribution functions.
From \cite{KS1912}.}
\label{fig:KSmath}
\end{figure}

The first concerns Kapteyn's {\it Star Streams}. Not long after Kapteyn's
presentation of these in 1904, Schwarzschild noted that the interpretation of
two distinct streams of stars moving through one another was not the only
possibility. He had described the distribution of peculiar
velocities of stars in space as much like
that of molecules in a gas, which means  according to the Maxwell distribution.
These velocities then are distributed as a Gaussian curve, but isotropic, so
Gaussians with the same dispersion (or mean velocity) in all directions.
Karl Schwarzschild proposed a distribution that was also of this Maxwellian
form but with dispersions that were different in three different
perpendicular directions. This then
is in the observations equivalent to two opposite
streams in the direction of the largest dispersion. It became known as an
ellipsoidal distribution, which then would mimic two opposite
streams while in fact it was a single distribution that happened to be
anisotropic. Kapteyn has stuck to his interpretation of the distinct streams
on the basis of the observed property that the make-up of stars in both
streams was distinctly different. Prominent British astronomer
Eddington supported that view for a long
time as well, until the evidence for different compositions of the streams
disappeared with better data. Schwarzschild's proposal was adopted definitely
after Jan Oort had discovered Galactic rotation and from dynamics showed
that the long axis of the velocity ellipsoid had to point towards and away
from the center of the Galaxy. That Kapteyn's streams deviated from this by
$20^{\circ}$, an observational result that turned out to be correct and referred
to as the deviation of the vertex, remained a problem until it was explained
by Oort later as due to the gravitational influence of spiral structure. To
honor his work in this area the velocity ellipsoid remains to be referred to
as the {\it Schwarzschild distribution}.

\begin{figure}[t]
\begin{center}
\includegraphics[width=0.98\textwidth]{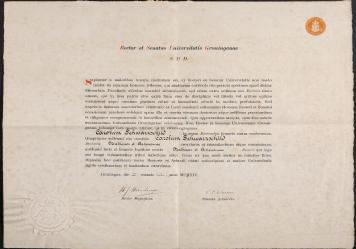}
\end{center}
\caption{ Diploma or `doctoral bull' for the honorary degree of Karl
Schwarzschild from the University of Groningen in 1914 \cite{Gott}}
\label{fig:bul}
\end{figure}

There was however another contribution of Schwarzschild that might have been the
strongest argument in favor of the doctorate {\it h.c.} and that has to do
with the derivation of the spatial distribution of stars from counts as a
function of apparent magnitude. Suppose that we know the distribution of
stellar types, and thus absolute magnitudes, at any position in the system.
This `luminosity function' was assumed to be the same everywhere and in
principle can be determined locally using statistical studies and secular
parallaxes. Now if we know this luminosity function and the stellar density,
the distribution of stars, the counts of stars in a particular direction as a
function of apparent magnitude, follows from a summation using these
two functions. Take an apparent magnitude and a position on the sky. For each
element of distance from the sun along the  corresponding line of sight each
apparent magnitude corresponds to a certain absolute magnitude. The number of
stars seen at that apparent magnitude from this element then is the total
density of stars there multiplied by the the fraction of stars of the required
absolute magnitude, which is the value of the luminosity curve at that absolute
magnitude. All such contributions from elements at other distances
along the line of sight,
corresponding to different absolute magnitudes, then have to be added up to
find the total count of stars of that particular apparent magnitude.

Mathematically this is an integral,
which can be evaluated in principle in a straightforward
manner. But the solution required is the inverse. Given the form
of the luminosity curve and the star counts at apparent magnitudes, the
problem is to determine the total density as a function of distance. That means
`inverting' the integral and inversion of integral equations
is notoriously difficult. Now in 1912 Karl Schwarzschild, being interested in
this problem, possibly since he was a student after all of von Seeliger, had
developed a method to solve this problem for the case that the
luminosity curve was a Gaussian function (see Fig.~{\large \ref{fig:KSmath}}).
Of course this was of enormous
importance to Kapteyn. Indeed, later in 1920 he and van Rhijn chose the
Schwarzschild method to solve for the distribution of stars in space, which
resulted in the Kapteyn Model.\footnote{ When the luminosity
function is unknown, there actually is a second
integral equation to be solved simultaneously and that involves mean
parallaxes as a function of apparent magnitude, derived from proper motions
and secular parallaxes. Kapteyn had studied this mathematical problem around
1900 with his brother Willem, who was a professor of mathematics in Utrecht.
This remained cumbersome and Schwazrschild's method was to be preferred.}
\bigskip

There are in the archives of the University of Groningen many notes related to
the honorary doctorates in 1914, but although discussions about
how to choose the candidates are recorded
in minutes of the meetings of the Senate and Faculties, there is no record of
supporting arguments for the choices. It seems obvious that Kapteyn had the
work described above in mind when he proposed Schwarzschild.
The archives in G\"ottingen
contain the letter to him in which he was notified of the award. In accordance
with the customs of the time it is in Latin. No explanation of for what work
the award was made, was given in the letter. Also the diploma or bull was
phrased in Latin (see Fig.~{\Large \ref{fig:bul}}). 

The question is whether or not Karl Schwarzschild actually came to Groningen to
receive the degree in person and to attend the celebrations of the tricentennial.
The answer is he did. The direct evidence consists of two small pieces of 
paper, on which he wrote rather short notes to the Rector Magnificus. They
were of course  in German, translated into English they say:

\noindent
\begin{quotation}
{\large Potsdam, 20-IV-1914.\\
To the Rector and Senate of the University of Groningen\\
I express my cordial thanks for the honor you have extended to me. It will be a
great pleasure to attend the celebration of the three hundredth anniversary of
the university.\\
Your sincerely dedicated\\
K. Schwarzschild.}
\end{quotation}

\noindent
\begin{quotation}
{\large Potsdam, 5 July 1914\\
To the academic Senate of the National University of Groningen,\\
I express cordial thanks for the sending of the beautiful medal that will help
me to keep alive  the remembrance of the impressive celebration in Groningen.\\
Your fully dedicated\\
K. Schwarzschild.}
\end{quotation}

The archives of the University of Groningen contain a printed note by the
Rector Magnificus, specifying that the procedure to come to the selection of
doctores {\it h.c.} would be that each Faculty (there were five) would
nominate six scientists for de honorary doctorates plus another fifteen
representatives of foreign universities and learned societies, for a total of
45. The Senate would take care that the final lists would be uniform in
composition of nationalities, disciplines, etc. There is no good record in 
the archives of the Senate, Rector Magnificus or Faculty of 
Natural Sciences of which professor proposed which candidate. It would be a
safe bet that Kapteyn had proposed Karl Schwarzschild.

The university had 43 professors in five Faculties: Theology (6), Law (6),
Medicine (11), Arts and Philosophy (10), and Natural Sciences
(10). It must have been too
difficult to decide which professor could and which could not propose a
person and have his proposal accepted,
so that -- in spite of the 45 proposed by the Rector -- in the end the number
grew to 67. 

\begin{figure}[t]
\begin{center}
\includegraphics[width=0.98\textwidth]{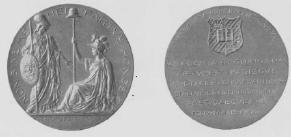}
\end{center}
\caption{ The memorial medal that the University of Groningen had had
struck for the celebration of its tricentennial in 1914. From \cite{RUGgedenk}.}
\label{fig:penning}
\end{figure}

The file in the Archives with organizational details of the tricentennial
contains long lists of guests and representatives to the proceedings, and
since the hotel accommodations in Groningen were limited, visitors were for
the larger part invited to stay with professors in their private homes. 
The archives contain a map of the city of Groningen so that visitors could
find their way around. It was oriented with the {\it west} at the top (note that
the word orientation comes from putting the {\it east} at the top). The
accommodation lists show that Schwarzschild and another honorary doctor, Peter
van Anrooy, had been staying with Kapteyn. No doubt Karl Schwarzschild came to
Groningen to receive the honorary degree and attend the celebrations.  He is
also on the printed list of the 283 attendees of the celebratory dinner, as
are Kapteyn and van Anrooy. 

The celebrations started on Monday June 29, with a session where guests were
received. A special commemorative book was presented \cite{RUGgedenk}; it
included an over 200 page history of the university written by well-known
professor of history Johan Huizinga (he would move to Leiden in 1915) and
various other contributions on the university in the past or present,
including a description by Kapteyn of his Astronomical Laboratory. Also a
special memorial medal was presented (Fig.~{\Large \ref{fig:penning}}), of
which Schwarzschild later had received a specimen according to his note above.
The formalities of the celebration took place on June 30 in the Nieuwe
Kerk, a major church in Groningen, but not in the large Martini Church in the
center of town where currently such festivities and ceremonies take place. The
program wasfull of speeches by dignitaries and representatives from
other universities and learned societies. Kapteyn had been asked by the US
National Academy of Sciences to be its formal representative and given a text to
read out (see \cite{JCKbiog} for details); he had been elected a Fellow of
the Academy in 1907. On the following day, July 1,  a special session of the
Senate took place during which the honorary degrees were officially bestowed
upon the 67 laureates. The Rector Magnificus took care of the one for H.M.
Queen Wilhelmina, the others were granted by the Deans of the Faculties, so
there was no specific individual promotor. For Schwarzschild the Dean was
biologist Jan Willem Moll. The degrees were bestowed by pronouncing the
official formula in Latin. Karl Schwarzschild was honored for his `exceptional
contributions to many aspects of astronomy and physics', and was proclaimed
-- somewhat contradictory -- doctor in mathematics and astronomy.
\bigskip

Nowadays Karl Schwarzschild is principally known for the Schwarzschild radius of
a black hole. This was work he did later. Weeks after the tricentennial the
First World War broke out and Schwarzschild -- although already 40 years of
age -- volunteered for the German Army, in which he eventually rose to
lieutenant of the artillery. In 1915 he was stationed in Russia where he wrote
three articles on General Relativity and Quantum Mechanics. He solved Einstein's
field equations for the case of a point mass -- or a single piece  of very
concentrated matter -- and found that there was an event horizon. Gravity is so
strong that light cannot escape through this horizon so that an outside
observer cannot be aware of any events inside. This dimension of a black hole
(a term coined later), defined as that of the event horizon, became known as
the Schwarzschild radius.

In 1915 in Russia Schwarzschild began to suffer from pemphigus, a rare
autoimmune skin disease of which he died in 1916 at the age of only 42.
\vspace{1cm}

\noindent
{\Large {\bf Interlude: Dr.h.c. P.G. van Anrooy}}
\bigskip

\noindent
Peter Gijsbert van Anrooy (1879--1954), who also stayed with the Kapteyns,
was a musician, composer and conductor. He had been director of the symphony
orchestra at Groningen (the Groninger Orkest Vereeniging) between 1905 and
1911; later he worked in the Hague as director of the renowned
‘Residentie-orchestra’ (Den Haag is the place of residence of the government)
and ‘Toonkunst-choir’ (toonkunst is a now somewhat out of date word for music
in general). In 1914 he worked in Arnhem in between these two assignments.

Henriette Hertzsprung-Kapteyn wrote a biography of her father \cite{HHKEng},
from which I quote (in my English translation, also available on
\cite{HHKEng}):
\begin{quotation}
{\large
‘It was around his 60-th birthday that he started to attend a course about
music in order to better appreciate this art. He was not gifted as a musician,
did not play an instrument and did not have a good singing voice. He had his
own manner of expressing himself using music, which at home was referred to as
‘trumpeting’. As soon as he started producing this sound, Mrs. Kapteyn could
not resist going to the piano and accompanying him, which had a very original
effect. You could hear him coming home while singing and sometimes while
working he would suddenly start singing, usually parts of sonatas or
symphonies that he was familiar with. He wanted to hear the same old pieces
again and again, the new unknown ones having little appeal to him. But the art
in itself, the depth of it, was a mystery to him; it filled him with a quiet,
respectful awe and for an artist he felt the deepest admiration. He followed a
course, given by Peter van Anrooy, at that time the conductor of the Groningen
orchestra, which interested him enormously. Every Wednesday evening he attended
the concert of the orchestra in the ‘Harmonie’, concentrating on the beauty of
music and always felt enriched by it. His acquaintance with van Anrooy, which
soon became a close friendship, took him closer to art. He found many parallels
between science and art. Isn’t it true that both are in their ideal form
unselfish and striving towards truth and purest expression? Oblivious to
earthly fame and prosperity in order to give the highest that a person has to
give? In that way he regarded art as the sister of science.’}
\end{quotation}

At the funeral of Kapteyn, van Anrooy played on the organ the final choir from
Bach’s {\it Matth\"aus Passion}. Considering this friendship and teacher-pupil
relation it is not unlikely that van Anrooy was actually also proposed by
Kapteyn for a doctorate {\it h.c.} and was subsequently asked to put him up
during the festivities. The Dean of the Faculty of Arts and Philosophy,
professor in English literature and Sanskrit, Johan Hendrik
Kern, praised van Anrooy in his laudatory as `expert musician, excellent and
skillful orchestra conductor, who in our country successfully raised music
to a higher standard.' He was promoted to doctor of arts.

I note before ending this section that Andrew Carnegie, who I suspect had
been proposed by Kapteyn,  received an honorary
doctorate in the Faculty of Law for ‘his dedication to the laws of war and
peace, not only in words but particularly in deeds’.

\section{ Dr.h.c. A.J. Cannon and Kapteyn's retirement
  in 1921}
Annie Jump Cannon (1863--1941) (see Fig.~{\Large \ref{fig:AJC}})
studied physics from 1880 to 1884
at Wellesley College in Massachusetts, which was (and is) one of the top
academic schools for women in the USA.
Her middle name `Jump' was her mother's maiden name.
She returned home, developing an interest in  photography
and studied all aspects of it, perfecting her skills and as a photographer
she traveled through Europe. In 1893 she published her photographs
after returning home which drew some attention. At about the same time she
suffered from scarlet fever which left her almost deaf.
After her mother's death, she returned to
Wellesley where she was hired to teach physics. It was also there that
she developed an interest in astronomy and spectroscopy. She studied astronomy
at Wellesley and for this took courses at Radcliffe College not far from
Harvard College, where Edward Pickering, who taught there, noted her and
hired her in 1896 as one of his assistants. She finished her Masters at
Wellesley in astronomy in 1907.
At Harvard, Cannon became a member of the Harvard
group of female computers that Pickering had  hired.

\begin{figure}[t]
\sidecaption[t]
\includegraphics[width=0.64\textwidth]{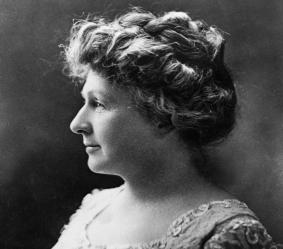}
\caption{ Annie Jump Cannon in 1922. From Wikimedia \cite{AJC22},
copyright expired.}
\label{fig:AJC}
\end{figure}

When the first spectra of stars had been obtained it became quickly clear that
there was a large variety among them. The usual attack then is to design a
classification scheme to order the characteristics. The resulting sequence
then may be, but does of course not have to be,
interpreted as a evolutionary one. This happened with
stellar spectra, giving rise to terms `early' and `late', while in fact
the sequence is to a large extent one of mass.

\begin{figure}[t]
\begin{center}
\includegraphics[width=0.80\textwidth]{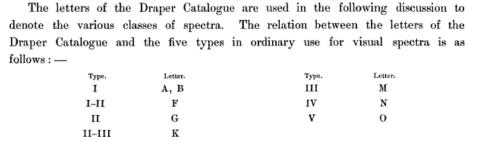}
\includegraphics[width=0.80\textwidth]{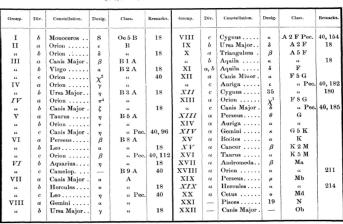}
\end{center}
\caption{ Reproduction of the original introduction of the spectral
types as we know them today in 1901. At the top the proposed order (`N' later
dropped from being part of the sequence and `O' was moved to the front). Below
some typical stars and their classifications. From \cite{AC1901}.}
\label{fig:ACtypes}
\end{figure}

Stellar classification started by the work of Pietro Angelo Secchi SJ of the
Specola Vaticana, who in the 1860s examined spectra of a
few thousand stars. He divided them into five classes, of which classes I to III
are more or less what we now would call `early', `middle', and `late'. The
current scheme is defined in the {\it Henry Draper Catalogue}. 
Henry Draper was a physician, but also an accomplished
amateur astronomer. Draper was one of the first to photographically record
the spectrum of a star at his private observatory and recognized the 
similarities in the many stellar spectra he obtained
to that of the sun. After his death his widow funded the
spectroscopic work of Pickering at Harvard through a fund called
the `Henry Draper Memorial'. Pickering obtained
photographic spectra of hundreds of stars 
and these were classified, in first instance  by Williamina Fleming into a
new system, splitting up Secchi's type I in types A through D, 
type II in E through L, designating type III as M and introducing a new type O
(from `Orion' since some stars in that constellation belonged to this type).
The result was published as the {\it Draper Catalogue of Stellar Spectra}
in 1890. This was refined by Antonia Caetana de Paiva Pereira
Maury, but particularly by Annie Cannon (using also spectra from the southern
station of Harvard College Observatory in Peru), who defined the now
universally adopted sequence O, B, A, F, G, K, M, each subdivided into ten
sub-classes. She is told to have been able eventually to classify 200
stars per hour or 18 seconds per star. This resulted in the
{\it Henry Draper Catalogue}, published by Cannon and Pickering in nine
volumes of the Harvard Annals between 1918 and 1924, containing about
250,000 stars (and later extended with the {\it Henri Draper Extension}). 

Fig.~{\large \ref{fig:ACtypes}} shows two tables from the original publication
in 1901, in which Cannon proposed more or less the sequence as we know it now
and its subdivisions (although the second capital letter later disappeared).
The type `N` was later considered not part of the sequence and the `O' was moved
to the top of it.

The proposal by Kapteyn to award Cannon an honorary doctorate is conserved in
the Groningen Archives. To preserve the authenticity I will reproduce it in
translation in full.
\begin{quotation}
{\large `It is known that an entirely new field of research in classical
astronomy has been added by the ascent of spectroscopy. A field in which
insights have been given that in the past seemed impossible. -- What we know
about the evolution of cosmic bodies is based entirely, or at least for a
major part, on spectroscopic observations and with the researches in the last
few years it appears that the role spectroscopy will play in the study of the
structure and motions of the sidereal world will become an almost equally
fundamental one.

In the meantime the faintness of stars are raising obstacles for the study, 
obstacles that for the moment can be overcome only for the brightest stars
and then only after inexhaustible patience and tireless energy.
Science however has to no lesser extent a need for spectra of fainter stars. It
is especially for the faint stars, stars of magnitude 12, that statistical
investigations have led to results concerning the structure of the universe.
It is reasonable to expect that when such investigations can be done separately
for each of the spectral types that the harvest of important results will be
much richer.

Unfortunately we are not there yet, but the most important step in this area
appears to have been taken. After the Observatory in Potsdam in 1883 had
obtained for a part of the sky spectra for 4000 stars, nothing was done in
this field that in importance is a match to the work at the observatory at
Harvard. In 1890 the so-called {\it Draper Catalogue} was published for all
stars that can just be seen by a sharp, naked eye. In total 10,000 stars.

Until almost the present time this is all that astronomers in general terms
had at their disposal. The prospect of much extended determinations was not
good. Telescopic stars need much more work and their number grows disturbingly
quickly with every next unit of magnitude.
Ten years ago no astronomer would have expected that now in 1921 we would have
available 200,000 carefully determined spectral types of nearly all stars up
to the $9^{th}$ magnitude plus a large number even beyond that of $10^{th}$ or
even the $11^{th}$ magnitude. A quarter of those, comprising four tomes is
already available to astronomers, the rest is in press and will presumably be
accessible to everybody within a year.

This gigantic effort, a true monument of organization, skill and perseverance
is due to a single woman
    \begin{center}
      \underline{Annie J. Cannon.}
    \end{center}
For all those who wish to study the evolution and the structure of the great
stellar system the appearance of this publication is the most important event
in the last few years. It opens the possibility described above to study
separately stars of different spectral classes, classes just as diverse as the
classes in the animal kingdom. What is missing to reach the same limit for
all stars will also not take too long because completeness for the
faintest stars will not be required.
    
In what urgent need the work of Miss Cannon will provide, is most clearly
illustrated by the fact that the Harvard Observatory  as reaction to urgent
insistence of a large number of astronomers has made, by mail, available the
manuscript with an enormous mass of data.
The Groningen Laboratory would not have been able to complete its latest
publications without the thousands of spectra that Miss Cannon with
extraordinary kindness has made available.

I believe that it is a real obligation for astronomers to give a proof of
their gratitude and bring homage for a piece of work for the satisfactory
completion of which except for Miss Cannon possibly no one else in the world
would have had the skill, perseverance and self-sacrifice; a piece of work so
urgent and of such far-reaching significance.
And since Groningen, possibly more than any other university, had the fortune
to profit from this work, I believe that the Senate and University of Groningen
should not forgo the privilege to associate itself with the author
    \begin{center}
      Miss Annie J. Cannon
      \end{center}
by offering her an honorary doctorate in Mathematics and Astronomy.'}
\end{quotation}

The Groningen work Kapteyn refers to is that of van Rhijn, who was making
preparations, by determining mean parallaxes as a function of apparent
magnitude, to repeat the analysis for the distribution of stars in
space separately for various spectral types.
The proposal was approved and Annie Cannon was duly informed of the decision.
No record of this is present in the Groningen Archives, nor among the scans
that were made for me of the Cannon Archives files on the Groningen and Oxford
honorary doctorates at the Harvard University Archives -- Pusey Library. The
letter in reply is available at Harvard in handwritten and typed form, the
latter being identical to the handwritten version that was actually sent to
the Groningen Rector Magnificus. In part it reads:
\begin{quotation}
{\large
`... To be ranked among the scholars of that ancient and renowned seat of
learning founded just before the Pilgrim Fathers left Holland for the
wilderness of my own Country, to be thus linked with the Groningen Astronomical
Laboratory, made famous by Professor Kapteyn, the world's greatest astronomer
of to-day, is indeed the highest honor of my life, far exceeding my fondest
dreams.

It will be impracticable for me to go to Groningen to receive the diploma in
person, and therefore I shall be most grateful to you if you will be kind
enough to forward it by mail. ...'}
\end{quotation}
So Cannon did not come to Groningen, which may have been a real
disappointment for Kapteyn. There was then no special celebratory session of the
Senate at which the doctorate would have been bestowed.
 The files on her
Oxford honorary doctorate in the Cannon Archives actually contains a letter
from Kapteyn concerning her Groningen degree, dated 10 June 1921:
\begin{quotation}
{\large `Dear Miss Cannon,\\
Let me be the first to congratulate you on the well earned honour
conferred on you by the University of Groningen. As far as I know you and
Schwarzschild are the only persons upon whom the doctorate in Mathematics and
Astronomy `honoris causa' has been bestowed. I hope you will find in this
tribute at least some small return for a work which even in Astronomy has
hardly a parallel, a work which is so urgently demanded for the further
progress of science and which will earn for you the gratitude of all who try
to penetrate somewhat further in the mysteries of the stellar Universe.\\
    \hspace{5cm} Very truly yours,\\
    \hspace{5cm} J. C. Kapteyn'}
\end{quotation}
Kapteyn fails to mention Easton, so maybe he did not consider him to be a real
astronomer. However the phrasing is such that Easton should have been mentioned.
The diploma was sent from Groningen on July 6, accompanied by a letter from
the Rector expressing regret she would not come but `understanding her
motives'. It is noteworthy that the files in the Cannon Archives do contain a
large number of congratulatory letters, telegrams and cards, mostly from
friends and organizations (such as the President of Wellesley College), but
only one from an astronomer (Otto J. Klotz, director of the Dominion Observatory
in Canada).

Among the further honors Cannon received were honorary doctorates from
Wellesley College, her Alma Mater, where she had been both a student and a
teacher, and from Oxford University in the UK. She had received only a
Master's degree form Wellesley but had not continued to submit a PhD thesis,
so this was a special honor. This was in 1925 and the two
proposed dates for the ceremonies almost excluded Annie Cannon from attending
both. Not much correspondence is available on the two degrees in the Cannon
Archives. The only relevant piece is a draft of a letter to the President of
Wellesley College, which begins as follows:
\begin{quotation}
{\large `It is with the deepest appreciation of the high honor proposed for me
by the trustees of Wellesley College that I write you my acceptance and
expectation to be present on May twenty-ninth. It is the very day I was
booked  to sail, but I find that I can change to a boat going a few days later
which will put me in England in due season. ...'}
\end{quotation}
There are essentially no congratulatory letters, telegrams, etc. in the
Cannon files associated with these two honorary degrees, so these must have
filed somewhere else.

There is a well-known photograph of Annie Cannon where she wears the gown
and cap associated with her doctorate {\it h.c.} from Oxford. She seems to have
hoped to receive such a gown also from the University of Groningen. Both the
Cannon and Groningen files contain a letter from the Rector Magnificus (original
in handwriting, copy in Groningen typed) in response to a (missing) letter
from Cannon. The reply letter is dated October 26:
\begin{quotation}
{\large `Dear Miss Cannon,\\
I have your letter of 10 October, for which my thanks. There is \underline{no}
gown or hood connected with the doctorate in this country; it was abolished
already more than a century ago. Only the professors of the universities wear
on official occasions a black velvet gown with white `chabot' and black velvet
cap. The universities have no colours nor utterly decorations, -- all in the
severe puritan style, as you see. I hope you will console yourself about this!'}
\end{quotation}  

This all is not to say Annie Cannon would not have appreciated the high honor
from Groningen. She obviously wore the Oxford gown and cap with pride. It is
fully understandable she would have been more taken by and would not fail
to attend the ceremony of the honorary doctorates
from her own Wellesley College and the renowned University of Oxford.

\section{ Discussion}
In the first place I briefly discuss Kapteyn and his honors.
He has been awarded three doctorates {\it h.c.} It is of interest
to note that there are none from universities in England, France, Germany,
Russia, etc., that were leading nations in astronomical research. This may not
be significant since Kapteyn's honors (listed in \cite{JCKbiog}, Appendix A.4)
show that he definitely received high honors from a number of these
nations, such as
knighthoods, medals and memberships of academies and learned societies. By the
way, the honorary degree from Edinburgh is unrelated to Kapteyn's collaboration
with Gill, who after all was Scottish; he was raised in Aberdeen and studied
there under Maxwell, and had nothing to do with Edinburgh.

It is interesting to briefly compare Kapteyn to his famous pupils de Sitter
and Oort. Willem de Sitter has been awarded four honorary doctorates: Cambridge
1925, Cape Town 1929, Wesleyan 1931 and Oxford 1932. Did he take the initiative
for Leiden University to award honorary degrees to astronomers? Yes, in three
cases he did, namely for Robert T.A. Innes in 1923 and Friedrich K\"ustner and
Henri-Alexandre Deslandres in 1928.\footnote{\normalsize
The site of Leiden University,
that lists honorary doctorates \cite{ULhc}, erroneously lists Deslandres as
having been awarded the doctorate in 1900.}
Innes was director of the Unie Sterrewag in Johannesburg, with which de Sitter
at about the same time had negotiated a collaboration and the establishment of
Leiden's southern station at its premises
in Johannesburg. K\"ustner was emeritus director of
the Sternwarte Bonn, and Deslandres was a French astronomer,
director of the l'Observatoire de Meudon et l'Observatoire de Paris.
The award was on the occasion of the Leiden
General Assembly of the International Astronomical Union, of which de Sitter
was president at the time. There is more to note about this; de Sitter wanted
to end the isolation of Germany that after the First World War had been
excluded from international organizations (see below for more on this)
and he used his prerogative as
President to invite individual German astronomers. In this spirit he took the
initiative to honor a French and a German astronomer  with a doctorate {\it
h.c.} He had some difficulty convincing the Leiden Senate to award two
honorary titles in one discipline at the same time. While K\"ustner worked in
astronomy relevant to the Leiden program -- he had continued the tradition of
the {\it Bonner Durchmusterung} by setting up photographic programs for
positional astronomy, in particular for proper motions --, Deslandres
did not, his field was
the sun and its atmosphere. He was the ranking French astronomer and had been
one of the vice-presidents of the IAU since 1922 and was ending his term in
Leiden.

More remarkable is Jan Oort, who had ten honorary
doctorates, including Oxford, Cambridge and Harvard (see \cite{JHObiog},
Appendix A.5), yet has not taken even a single
initiative to award one by Leiden University.
In fact the next astronomer after K\"ustner in 1928  to receive a Leiden
doctorate {\it h.c.} is later Nobel Prize laureate Reinhard Genzel in
2010!\footnote{ In a
formal sense this is incorrect, since Jacob Evert Baron de Vos van Steenwijk
received one in 1959 in the Faculty of Law. He held a PhD in astronomy (from
1918 under Ernst van de Sande Bakhuyzen), but became a politician and
administrator, with appointments such as Mayor, Queen's Commissioner and
President-Curator of Leiden University. But throughout his career he has been
actively involved in organizational bodies concerning Dutch astronomy, and
attended IAU Genral Assemblies.}
Pieter van Rhijn did not receive any honorary doctorates and did not take the intiative to have the university of Groningen award an other one. In fact no astronomer received a doctorate {\it h.c.} from Groningren after Cannon.
\bigskip

What about Kapteyn's choices for Schwarzschild and Cannon? Who else might
he have considered? Some of the most
critically important persons for him in establishing his career and
international fame had been David Gill (for the {\it Cape Photographic
  Durchmusterung}), Simon Newcomb (for making his work known in America and
his native Canada, and inviting him in 1904 to the St. Louis Congress), George
Hale (for inviting him to Mount Wilson as Carnegie research associate and
adopting the {\it Plan of Selected Areas} as prime observing project for the
60-inch) and Edward Pickering (for a major contribution from
Harvard to the {\it Plan}).

Gill and Newcomb had died by 1914, but Pickering lived until 1919 and would have
been an option in 1914. Kapteyn had received his honorary doctorate from Harvard
in 1909 and could have reciprocated the honor. Kapteyn was very sensitive about
giving credit where it belonged and realized that the work for the {\it Henry
Draper Catalogue} must have been done really for the most part by Annie Cannon.
There had been disagreements with Pickering, especially about the latter’s
insistence to include a {\it Special Plan} of areas in the Milky Way, which
meant much
extra work; it had annoyed Kapteyn that he had to accept this to ensure
Pickering’s cooperation. In the end he would have considered
Schwarzschild’s work more suitable and had better relations with him.
In any case, from a scientific as well as personal
point of view Schwarzschild probably seemed a more agreeable choice.

Hale certainly would have been a possibility in 1914. This might have resulted
in an uncomfortable situation with Andrew Carnegie receiving the same honor
at the same time, who after all in a sense
was Hale's superior or `boss'. In fact, it is far from
ruled out that Kapteyn had been involved in selecting Carnegie (who else
on the Groningen university faculty would have a connection to him?) and then
also including Hale in addition might have been too much. 

Then there is Anders S. Donner, who Kapteyn had first met at the great {\it
Carte du Ciel} congress in Paris in 1887. Donner was director of the
Helsingfors Observatory (Swedish for Helsinki), where he acquired plates for
his participation in the {\it Carte du Ciel}. Kapteyn and Donner had quickly
become very good friends (they were among the younger attendants in Paris and
not part of the establishment), and over the years Donner had collected an
enormous amount of photographic material for Kapteyn, which resulted in quite a
number of joint papers in the {\it Publications of the
Astronomical Laboratory at Groningen}. Until 1925, 39 volumes
appeared in these {\it Publications},
of which no fewer than nine were based on plates taken by Donner.
However, Donner had not published himself any significant papers; his vocation
seemed to have been operating the telescope (at least in winter;
at his latitude the observatory was closed from April through August
when at night there is no end to astronomical twilight) and providing
photographic plates for others.

\begin{figure}[t]
\sidecaption[t]
\includegraphics[width=0.64\textwidth]{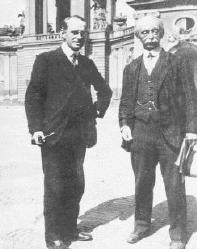}
\caption{ Arthur Eddington and Jacobus Kapteyn. It is not known
when, where and by whom this photograph has been taken.
Oort was presented with a copy  
of this by S. Chandrasekhar. For the full story see \cite{JHObiog}.
It would seem that this photograph has been taken around 1920. 
From the Oort Archives.}
\label{fig:EddKap}
\end{figure}

There is one other person who has meant much to Kapteyn,
and who might certainly have
qualified for an honorary degree and that is Arthur S. Eddington
 (see Fig.~{\Large \ref{fig:EddKap}}).
Without doubt Eddington was one of the most prominent astronomers of the
twentieth century. When Kapteyn had presented his {\it Star Streams} in 1904,
within two years Eddington,
who had just become chief assistant at the Royal 
Greenwich Observatory after having been one of the brightest students in 
Cambridge,
devised a quantitative test and applied this to a set of stars in the
{\it Groombridge Catalog}
for which new proper motions were available at Greenwich.
Where Kapteyn used brighter stars 
all over the sky, Eddington's material constituted more stars over a limited
part of the sky, which also were fainter. 
His results did, as he wrote, `strongly support Kapteyn's hypothesis of two
star-drifts'.

Now in 1914, Eddington was only 32 years of age and this would seem a bit young
for a doctorate {\it h.c.} But in 1921 he certainly could have been a good
candidate. In addition to confirming the {\it Star Streams}, in his
book {\it Stellar Movements and the Structure of the Universe}
from 1914, which had drawn much attention,
Eddington had prominently discussed the concept. In his
introductory address on the occasion of the centenary celebration of the
RAS in 1922 he would list the in his opinion six `outstanding landmarks in
these hundred years' and Kapteyn's {\it Star Streams} was one of these.
He remarked: `But I think the great impetus on sidereal
astronomy  came from Kapteyn's discovery'. And he wrote later an obituary
of Kapteyn of unusually high praise. This was after 1921 of course, but he
must have had these opinions before that and been a promoter of Kapteyn's work
throughout the years. 

The selection in 1921 at the time of Kapteyn's seventieth birthday and
retirement has to be seen as well in the light of another development. Kapteyn
had alienated a significant number of leading astronomers in particular in
England (and maybe other parts of the UK as well), France, Belgium and the
USA, because of his stance after Wold War I on international organizations.
Especially in these countries there was a strong sentiment and
movement to ban the defeated
nations from such organizations that were being formed
just after the end of WWI. Kapteyn had vehemently opposed
setting up these without participation from Germany, Austria,
and the other Central Powers, in particular the
International Research Council and the International Astronomical Union.
More background on this can be found in an authoritative book by Kevles,
{\it The Physicists} \cite{Kevles}.
Together with his friend, physiology and  psychology professor Gerardus
Heymans, 
Kapteyn took the initiative to circulate an open letter strongly protesting
and condemning this attitude, that
in the end was signed by almost 300 persons (however Kevles fails to emntion
the role of Heymans; see \cite{JCKbiog} for an excerpt from this letter).

When Kapteyn's 70th birthday
was approaching in 1921, de Sitter took the 
initiative to prepare the publication of Kapteyn's selected 
works.  A committee that was formed quickly ran into opposition in the UK,
apparently because German K\"ustner was a member. So, when
Frank W. Dyson, who then was at Greenwich,
solicited help among British astronomers, he did get support from Arthur
Eddington, but strong opposition from many others as well. French astronomer
Jules Baillaud, Belgian George Lecointe, and English astronomer
Herbert H.Turner from
Oxford were among the most outspoken among European astronomers in
arguing for exclusion of German scientists after the war.  
They and others  incorrectly accused Kapteyn of having accepted the {\it
Orden pour le M\'erite} from the German Kaiser at the same time as the
captain of the submarine that had sunk the Lusitania.
The {\it Orden pour le Mérite} was a German distinction bestowed by the Kaiser.
Note that {\it Orden} is German. It was a
very high honor for a foreign scientist to receive that distinction (a few
foreign astronomers had preceded Kapteyn, such as Simon Newcomb, David Gill
and Edward Pickering, as well as German astronomers Freidrich Argelander and
Arthur von Auwers, but also famous scientists  had received it, such as
Charles Darwin, Lord Rayleigh, Hendrik Lorentz,and others). Kapteyn had
received the {\it Orden} in 1914 just before the War, at the same time as von
Seeliger and Max Planck. This was in the Scientific Class, but there also was
a Military Class, which the captain of the submarine that sunk the passenger
ocean liner {\it Lusitania} -- an act that enraged the world and played an
important role in the US joining the War --, had received, but that was in 1917!
Yet his stance did not prevent Kapteyn from  being appointed not long after the
War a corresponding member of the French Acad\'emie and foreign member of the
UK Royal Society. De Sitter's plan
came to naught and he and Kapteyn's successor in Groningen Pieter van Rhijn
might have resorted to a plan to have a doctor {\it h.c.} selected by Kapteyn.

Now Eddington was the first Englishman, who was prepared to work towards
reconciliation. 
George Hale seemed to have sided with the ones opposing Germany, but he
in any case defended Kapteyn as being far from pro-German. After all, in their
correspondence during the War, Kapteyn had expressed repeatedly his view that
Germany was to blame for the
War and had also very strongly condemned the sinking of the Lusitania.
In 1920 Eddington was the only Englishman that attended the congress of the
Astronomische Gesellschaft. Whether avoiding further escalation played a role
in not selecting
Eddington in 1921 remains a matter of speculation, but considerations like
these must have played a role, regardless of how appropriate it had been to
honor him this way on scientific grounds.

In 1922 an argument
against selecting Hale for an honorary doctorate might have been Kapteyn's
very poor relations with Walter Adams, who was Hale's deputy. Adams might have
seen this as a provocation against him, which Kapteyn would have wanted to
avoid.

The choice of Annie Cannon seems to have had, in addition to this,  two
other aspects. In the first place a
genuine appreciation and
admiration for the work of stellar spectral classification at Harvard
as a very fundamental and vital contribution to astronomy.  Cannon in that case
would easily qualify as the obvious first exponent of this work,
as Kapteyn expressed in his arguments to the Senate when he proposed her.
Kapteyn was rather strict about giving the credit to where it belonged as is
clear from the Adams-Kohlsch\"utter case. Of
course the {\it Henry Draper Catalogue} was compiled under Pickering's
directorship, but with the latter dead the honorary degree could easily go to
Cannon, where Kapteyn must have felt the credit belonged, without passing over
Pickering. The second thing is that what might have played also a role is
that Kapteyn would have welcomed the
fact that it then would be a woman, who would be honored. Kapteyn has a record
of supporting women's rights.

One of his best friends in Groningen was the well-known professor of
philosophy and psychology Gerardus Heymans (see \cite{JCKbiog} or
\cite{JCKEng}). Heymans was among other matters interested
in differences between men and women. He
researched this with questionnaires, asking professors and lecturers at
Dutch universities about character traits such
as individuality, ability for abstraction, memory, etc., and whether they
felt predominantly male or female. Kapteyn and his calculators helped reduce
these data and maybe contributed to this as
well. Although Kapteyn and Heymans were good friends, they often had different
opinions, but when it came to women's suffrage, they were in perfect
agreement. It is true that Heymans lobbied more actively for women's suffrage,
but according to a thorough study concerning {\it New participants in science:
Female students and teachers at the University of Groningen, 1871–1919}
\cite{IngedW},
Kapteyn was certainly involved in Heymans' circle of acquaintances who worked
for women's emancipation.

It is
also remarkable that Kapteyn was the first astronomer in the Netherlands with
a female PhD student, Etine Imke Smid. She obtained her PhD degree in 1914
on a thesis, which concerned the study of proper
motions of more than one hundred stars. She was the first woman in
the Netherlands to obtain a doctorate on an astronomical dissertation. After
her defense, Etine Smid worked for Kamerlingh
Onnes in Leiden for some time, but moved to Deventer after she got married and
left science.

And I quote from Henriette Hertzsprung-Kapteyn's biography:
\begin{quotation}
{\large `The children had grown up by now and all three of them were doing
academic studies. The two girls were among the first female students.
The oldest had chosen medicine and I took law as subjects, which gave
rise to much criticism in these turbulent days of the fight for women's
rights, but Kapteyn felt that females studying at universities was so
natural and unquestionable, that one did not get far with counter
arguments. The son went to Freiburg in Saxony to study mining
engineering.'}
\end{quotation}
The eldest daughter indeed 
studied medicine in Groningen, the younger law, for a while in Groningen,
but she later switched to English in Amsterdam. The son went to Freiburg to
study mining engineering; the fact that he did not go to Delft for this had
to do with the lower costs of studying in Freiburg. His choice to send his
daughters to university meant Kapteyn had to support three children during
academic studies and that was a major financial burden. Professors did earn a
decent salary, but many of them, including Kapteyn, did extra teaching in
order to be able to finance their children's studies.
Kapteyn chose to carry the financial burden of his daughters entering
university, even if this meant that his son would have to go and study
in Germany. Female students at universities were still quite rare; the first
female student at a Dutch
university (that of Groningen) had been Aletta Henri\"ette
Jacobs, who entered the university to study medicine in 1871, followed six
years later by her sister Charlotte who studied pharmacy,
hardly a generation before the Kapteyn daughters entered university.

So, Kapteyn was a supporter of women's emancipation. The
fact that he was an unconventional man in a number of ways may have
been a reaction to his strictly religious upbringing. But he did not allow
reform at all fronts. His marriage was along conventional lines; his wife
took care of the household, while Kapteyn saw the finances as his exclusive
responsibility.

Still, he had caught two birds with one stone: honoring the great progress in stellar
spectroscopy and selected a female doctor {\it h.c.} Excluding the avalanche
of honorary degrees in 1914, Cannon's was the first honorary doctorate
from Groningen to go to a
foreign scientist since Robert Koch in 1884. And the second female doctor {\it
  h.c.} from abroad
in the history of Groningen University after Alice Boole Stott in the
1914 tricentennial. Thos drought  would last until 1958, when Elizabeth Caroline Crosby,
American professor of neuroanatomy would become the third.

After his formal retirement Kapteyn left Groningen. He and his wife bought a
house in Hilversum and after some travel in Europe moved in with their daughter
and her husband in Amsterdam before settling in their new house. Leiden
Observatory offered him a parttime appointment to take charge of the
astrometric department. But them the first signs of what would prove to be his
fatal illness appeared and he died still in Amsterdam within a year of his
retirement. Henriette Hertzsprung-Kapteyn tells
us that Kapteyn and his wife left Groningen quietly. They had said farewell to
their friends during dinners in small numbers and left unnoticed one day at
seven in the morning. It is a pity that Annie
Cannon could not come to Groningen to accept the honor in person during a
special ceremonial meeting of the Senate to mark the departure of possibly
Groningen's most successful and celebrated professor.

﻿

\section{ Summary and conclusion}
Kapteyn at three occasions proposed doctores {\it honoris causa} for the
University of Groningen. The first one, in 1903, coincided with his
25-th anniversary as a professor in Groningen and went to Cornelis Easton. 
Easton was a journalist and amateur astronomer. The degree was awarded to
honor him for his work of mapping of the Milky Way by making drawings.
This he followed up by efforts to derive from this the spiral structure of the
Galaxy.

In 1914 he proposed Karl Schwarzschild particularly for his theoretical
contributions to the study of the structure and kinematics of the stellar
distribution in space.

Kapteyn most likely also proposed musician, composer and teacher Peter van
Anrooy, who not only developed his taste and appreciation for music but taught
him lessons as well to better understand music.

It is very well possible that he also had a determining role in the selection
of industrialists and philanthropist Andrew Carnegie, the benefactor of the
Institution that bore his name and provided Kapteyn’s position as research
associate, which facilitated much of his research including  his annual visits
to Mount Wilson Observatory. 

Maybe for this reason the option to select George E. Hale was not favored, while the
otherwise real option of Arthur S. Eddington was put aside because of his
relatively young age.

There were a few controversies that may have limited the choice in 1921 and
which have played an adverse role in many aspects of Kapteyn’s legacy. One was
the rather poor relation with Walter S. Adams from Mount Wilson Observatory,
which prevented his return to California after the War and which might have
escalated were Hale selected.

Another was Kapteyn’s stand on the in his view extremely objectionable policy
of excluding defeated nations as Germany, Austria, etc. from membership
of international political and scientific organizations. Eddington also very
strongly defended the same position as Kapteyn, but selecting him for an
honorary degree might have escalated tensions with others.

When Kapteyn retired in 1921 he proposed Annie Cannon for her fundamental
and essential contributions to the Harvard spectral classification and the
{\it Henry Draper Catalogue}. He might have felt that in this way she would
receive the credit she deserved, which for women would naturally go to the
male director under whom she worked. The fact that Edward Pickering had died
helped, since it meant that he did not have to be passed over.
An important argument for the selection of her  very
likely was that Kapteyn supported women’s equality and rights and welcomed the
idea that she would be the first foreign female thus honored prominently
(i.e. not as part of a large contingent) by the University of Groningen.
\bigskip

The imminence of the outbreak of the War, his enlisting in the German Army and
death during the War is probably the reason Schwarzschild’s honorary title is
seldom or not at all quoted, certainly not in obituaries. Cannon’s degree
is usually mentioned but is overshadowed by the one from Oxford. Her
decision not to come to Groningen meant no formal ceremony to mark Kapteyn’s
retirement and departure from Groningen. 
\bigskip

\noindent
{\bf Acknowledgements} I am grateful to the staff of the Archives at
Nieders\"achsische Staats- und Universit\"atsbibliothek G\"ottingen  and of
the Harvard University Archives -- Pusey Library for providing scans of the
relevant correspondence in the Schwarzschild and Cannon archives. I thank
historians Klaas van Berkel and David Baneke for a critical reading of
a draft of this paper and making very useful comments and suggestions.
I thank the staff of the Kapteyn Astronomical Institute for support and help
and for hospitality extended to an emeritus professor as guest scientist. 

}

\end{document}